\documentclass[11pt]{article}
\usepackage{fleqn,cospar}
\usepackage[dvips]{epsfig}
\usepackage{graphicx}

\newcommand{\ts}{\thinspace}
\newcommand{\cl}{\centerline}
\newcommand{\Z}{\sf Z}
\newcommand{\gtsim}{\lower.5ex\hbox{$\; \buildrel > \over \sim \;$}}
\newcommand{\ltsim}{\lower.5ex\hbox{$\; \buildrel < \over \sim \;$}}

\title{HEXTE STUDIES OF SCO X--1 SPECTRA: \\ 
       DETECTIONS OF HARD X--RAY TAILS   \\
       BEYOND 200 keV}

\author{F. D'Amico\address{UCSD/CASS, 9500 Gilman Dr., La Jolla, CA 92093-0424, USA}$^,$
                  \address{INPE, Av. dos Astronautas 1758, 12227--010 S. J. dos Campos - SP, Brazil},
        W. A. Heindl$^1$, R. E. Rothschild$^1$, L. E. Peterson$^1$, \\
        D. E. Gruber$^1$, M. Pelling$^1$, and J. A. Tomsick$^1$}

\begin{document}

\maketitle

\begin{abstract}
Using the HEXTE experiment on-board the {\it RXTE} satellite, we performed
a search for hard X--ray tails in Sco X--1 spectra. We found strong evidence
for the presence of such a  non--thermal component on several occasions.
Using the PCA/{\it RXTE} we were able to track the position of the
source along the {\Z} diagram, and we observed that
the presence of the hard X--ray tail is not confined to a particular
region. However, we found a correlation between the power law
index of the non--thermal component and the position of the source in
the {\Z} diagram, suggesting that the hard X--ray spectrum (i.e., $E
> 50${\ts}keV) becomes flatter as the mass accretion 
rate increases. We were also able to study the temporal variation of the 
appearance/absence of the hard X--ray component. With our derived 
luminosities, we were also able to test the idea that X--ray luminosities 
can be used to distinguish between X--ray binary systems containing neutron 
stars and black holes. 
\end{abstract}

\section{INTRODUCTION}

It is well established, after the {\it GRANAT}/SIGMA and {\it
CGRO} missions, that a power--law spectra extending beyond
30{\ts}keV are not exclusive signatures of a black hole in an X--ray
binary system (Barret and Vedrenne, 1994; Barret, McClintock, and
Grindlay, 1996). High sensitivity observations with {\it RXTE}
(e.g., Heindl and Smith, 1998; Barret et al., 2000) have shown the
presence of hard X--ray flux in other types of low--mass
X--ray binaries (LMXBs) and {\it
BeppoSAX} observations also have detected hard X--ray flux from LMXBs systems
(see, e.g., Frontera et al., 2000; Iaria et al., 2000; Di Salvo et al., 2000).

In this work we studied the hard X--ray spectrum of Sco X--1, a high
luminosity LMXB {\Z} source, using the High Energy X--ray Timing
Experiment (HEXTE) on--board {\it RXTE}. The presence of a non--thermal
component in Sco X--1 spectra has been historically reported (e.g.,
Peterson and Jacobson, 1966; Haymes et al., 1972; Duldig et al., 1983),
but much more sensitive searches have failed to detect such a component,
placing strong upper limits on the non--thermal flux (e.g., Greenhill et
al. 1979; Rothschild et al., 1980; Soong and Rothschild, 1983). Recently, 
Strickman and Barret (2000) reported the presence of a hard X--ray tail 
in Sco X--1 using {\it CGRO}/OSSE. Thanks to the HEXTE sensitivity,
we show here, on several occasions, strong evidence for the presence of
this variable non--thermal component extending  up to 220{\ts}keV.

This present study is an extension of a previous work (D'Amico et al.,
2000) with the use of an expanded database from the public {\it RXTE} archive of
Sco X--1 observations, containing data from 1997 April to 1999 July.
Our on--source time (for the Proportional Counter Array, PCA) is  
203,440{\ts}s in 22 subsets of data (an
improvement factor of 30{\ts}{\%} over our previous study) which,
in turn, provided us with more than 100{\ts}ks of on--source live time
in HEXTE. In the next sections we describe our data selection and
analysis, then we discuss the detection of a non--thermal component and
present our conclusions. 

\section{DATA SELECTION AND ANALYSIS}

We used data from HEXTE (Rothschild et al., 1998) to search for hard
X--ray tails in Sco X--1 in the $\sim${\ts}20--220{\ts}keV interval
and data from PCA (Jahoda et al., 1996) to determine the position of
the source in the {\Z} diagram. We have chosen, from the public {\it RXTE}
database of Sco X--1, those subsets of data in which
$\gtsim${\ts}5000{\ts}s of HEXTE total on--source time was available, in
order to achieve good sensitivity at high energies, resulting in 22
{\it RXTE} observations (as of 2000 August) that cover the period from 1997 April
to 1999 July. If, in a particular observation, the source was moving
back and forth from the normal to the flaring branch (see, e.g., van der Klis,
1995), then we split the data according to the branch resulting in the
28 observation segments displayed in Table 1.  Since we are primarily
interested in the study of the non--thermal ($\gtsim${\ts}50{\ts}keV) 
spectra of the source (the ``hard'' component of the spectrum), we have simply
modeled the (HEXTE) low energy thermal component as thermal bremsstrahlung
emission .

We refer the reader to D'Amico et al. (2000), for a more detailed 
description of the instrument and data analysis.

\begin{table}
\begin{center}
\cl{\bf Table 1.~~Selected RXTE Data Observations of Sco X--1}

\medskip
\begin{tabular}{c c c c c c c c c}
\hline
\hline
  OBSID           &  MJD  & No.& Z Pos.&  Live Time$^{a}$ & SNR$^{b}$ & F$^{c}$ & Flux$^{d}$ & Hard Tail (Y/N) \\
\hline
20053-01-01-00    & 50556 &  1 &  HB   &  6341  & 10.6  & 4.4E$-$15 & 8.57$^{+1.52}_{-1.38}$ & Y  \\
20053-01-01-02    & 50558 &  2 &  NB   &  5434  &  4.2  & 1.7E$-$7  & 4.05$^{+0.57}_{-0.87}$ &    \\
20053-01-01-03    & 50559 &  3 &  NB   &  5896  &  4.8  & 3.9E$-$9  & 5.20$^{+0.75}_{-0.74}$ &    \\
20053-01-01-05    & 50561 &  4 &  NB   &  1908  &  0.8  & 6.2E$-$5  & 6.29$^{+1.77}_{-1.77}$ & N  \\
20053-01-01-05    & 50561 &  5 &  FB   &  4593  &  2.4  & 7.7E$-$4  & 2.47$^{+1.94}_{-1.51}$ &    \\
20053-01-01-06    & 50562 &  6 &  NB   &  8558  &  7.3  & 3.5E$-$13 & 8.63$^{+0.91}_{-0.90}$ & Y  \\
10061-01-03-00    & 50815 &  7 &  NB   &  5484  &  4.2  & 2.7E$-$9  & 5.48$^{+0.69}_{-0.74}$ &    \\
20053-01-02-00    & 50816 &  8 &  FB   &  8145  & 18.6  & 5.8E$-$12 & 13.6$^{+2.0}_{-2.0}$   & Y  \\
20053-01-02-03    & 50819 &  9 &  NB   &  5686  &  3.5  & 2.7E$-$9  & 5.91$^{+0.96}_{-0.95}$ &    \\
30036-01-01-000   & 50820 & 10 &  FB   &  6858  &  5.8  & 3.8E$-$7  & 7.73$^{+1.62}_{-1.65}$ & Y  \\
20053-01-02-04    & 50820 & 11 &  FB   &  4193  &  4.9  & 3.3E$-$4  & 3.44$^{+1.23}_{-1.00}$ &    \\
30036-01-02-000   & 50821 & 12 &  NB   &  4370  &  2.2  & 3.6E$-$4  & 4.50$^{+0.87}_{-0.90}$ &    \\
30036-01-02-000   & 50821 & 13 &  FB   &  2527  &  1.8  & 3.6E$-$4  & 2.83$^{+1.05}_{-1.06}$ &    \\
30406-01-02-00    & 50872 & 14 &  NB   &  2326  &  2.5  & 1.0E$-$6  & 4.03$^{+1.07}_{-1.13}$ &    \\
30406-01-02-00    & 50872 & 15 &  FB   &  3125  & $<$ 0 & 0.08      & $<$ 2.20               & N  \\
30035-01-01-00    & 50963 & 16 &  HB   &  6253  &  6.2  & 1.2E$-$14 & 3.63$^{+0.92}_{-0.88}$ & Y  \\
30035-01-03-00    & 50965 & 17 &  NB   &  3750  &  1.3  & 2.4E$-$8  & 3.20$^{+0.86}_{-0.89}$ &    \\
30035-01-03-00    & 50965 & 18 &  FB   &  2318  & $<$ 0 & 0.03      & $<$ 2.67               & N  \\
30035-01-04-00    & 50966 & 19 &  FB   &  6230  & $<$ 0 & 0.38      & $<$ 1.00               & N  \\
40020-01-01-00    & 51186 & 20 &  NB   &  2669  &  6.6  & 1.2E$-$7  & 6.10$^{+2.21}_{-2.10}$ & Y  \\
40020-01-01-01    & 51186 & 21 &  NB   &  4233  &  3.5  & 3.3E$-$8  & 3.77$^{+1.66}_{-1.45}$ &    \\
40020-01-01-02    & 51187 & 22 &  FB   &  8000  &  8.1  & 2.2E$-$9  & 7.40$^{+2.46}_{-0.94}$ & Y  \\
40020-01-01-04    & 51188 & 23 &  NB   &  2836  &  1.8  & 1.1E$-$3  & 3.00$^{+1.28}_{-1.20}$ &    \\
40020-01-01-04    & 51188 & 24 &  FB   &  1822  &  1.8  & 0.11      & 2.95$^{+2.12}_{-2.48}$ &    \\
40020-01-01-05    & 51189 & 25 &  SA   &  4316  &  2.9  & 5.6E$-$4  & 2.57$^{+0.88}_{-0.87}$ &    \\
40020-01-03-00    & 51193 & 26 &  HB   &  3578  &  6.2  & 9.8E$-$10 & 8.45$^{+1.94}_{-1.69}$ & Y  \\
40706-01-01-000   & 51433 & 27 &  NB   &  8076  &  2.8  & 1.4E$-$5  & 2.09$^{+0.55}_{-0.53}$ &    \\
40706-01-01-000   & 51433 & 28 &  FB   &  2167  &  2.2  & 1.2E$-$3  & 2.45$^{+1.04}_{-1.07}$ &    \\ 
\hline
\hline
\end{tabular}
\end{center}
\smallskip
\noindent
Note: HB=Horizontal Branch, NB=Normal Branch, FB=Flaring Branch; 
\newline
\noindent
SA=Soft Apex ; No.=observation number

\smallskip
\noindent
$^a$ total live time for HEXTE, in s

\smallskip
\noindent
$^b$ signal to noise ratio in the 75-220 keV energy range

\smallskip
\noindent
$^c$ probability that a better fit occurred due to a random chance

\smallskip
\noindent
$^d$ flux in the 75--220{\ts}keV energy range, in units of 
    10$^{-10}${\ts}ergs{\ts}cm$^{-2}${\ts}s$^{-1}$; apart from the 5 detections
\newline
\noindent
of a hard X--ray tail, the photon index used for the non--thermal component was our average value 
\newline
\noindent
of $\sim${\ts}1; uncertainties are given at 90{\ts}{\%} confidence level
\end{table}

\section{RESULTS}

We developed two criteria to determine the presence of a hard X--ray
tail in a particular spectrum: 1) a signal to noise ratio (SNR)
$\geq${\ts}5 in the 75--220{\ts}keV energy range, and 2) an F--Test
null significance for the addition of the hard component at a level of
10$^{-7}$ or less. We claim that we observed a hard X--ray tail {\bf
only} when both of these criteria are fulfilled. We have 8 clear
detections of non--thermal flux in Sco X--1 spectra, with the hard
X--ray tail extending at least to 220{\ts}keV. The spectral parameters
derived for those observations are shown in Table 2. Similarly, we
defined a strong non--detection when 1) the SNR in the 75--220{\ts}keV
is $<${\ts}1, and 2) the F--Test was at a level of
$\gtsim${\ts}10$^{-5}$. We thus observed (see Table 1) 4 strong
non--detections. We note that several of our observations have very 
significant F--Test values, but with an SNR{\ts}$<${\ts5. In those 
cases we do not claim that we have observed a hard X--ray tail, 
although this is an open possibility. In Figure {\ref{fig1}} we show 
one of our clear detection spectra, together with a spectrum  when we 
have not detected a hard component.

\begin{table}
\begin{center}
\cl{\bf Table 2.~~Hard X--ray Tail Detections in Sco X--1}

\medskip
\begin{tabular}{c c c c c c c c c}
\hline
\hline
              & &\multicolumn{2}{c}{Bremsstrahlung}           &         &\multicolumn{2}{c}{Power--Law}                                                      & &    \\
\cline{3-4}
\cline{6-7}
    OBSID       & No. & kT (keV)              &        Flux$^{a}$      & &      $\Gamma$           &      Flux$^{b}$                 & $\chi^2_{\nu}$  & {\Z} Pos. \\
\hline
20053-01-01-00  & 1   & 4.83$^{+0.04}_{-0.05}$ & 9.03$^{+0.39}_{-0.36}$ & & ~1.75$^{+0.22}_{-0.20}$ & 1.56$^{+0.10}_{-0.11}$          & 1.59           &    HB     \\
20053-01-01-06  & 6   & 4.50$^{+0.05}_{-0.05}$ & 5.83$^{+0.30}_{-0.27}$ & & ~1.64$^{+0.29}_{-0.27}$ & 0.91$^{+0.11}_{-0.10}$          & 1.06           &    NB     \\
20053-01-02-00  & 8   & 4.36$^{+0.03}_{-0.03}$ & 7.16$^{+0.26}_{-0.24}$ & & -0.17$^{+0.30}_{-0.33}$ & 1.22$^{+0.13}_{-0.14}$          & 1.26           &    FB     \\
30036-01-01-000 & 10  & 4.28$^{+0.03}_{-0.04}$ & 9.63$^{+0.42}_{-0.41}$ & & -0.71$^{+0.63}_{-0.70}$ & 0.63$^{+0.11}_{-0.11}$          & 1.29           &    FB     \\
30035-01-01-00  & 16  & 4.51$^{+0.08}_{-0.07}$ & 7.45$^{+0.60}_{-0.67}$ & & ~2.37$^{+0.33}_{-0.28}$ & 1.04$^{+0.08}_{-0.08}$          & 1.09           &    HB     \\ 
40020-01-01-00  & 20  & 4.54$^{+0.06}_{-0.07}$ & 4.07$^{+0.24}_{-0.20}$ & & ~1.00$^{+0.53}_{-0.49}$ & 0.75$^{+0.16}_{-0.15}$          & 1.13           &    NB     \\
40020-01-01-02  & 22  & 4.41$^{+0.02}_{-0.02}$ & 7.46$^{+0.15}_{-0.15}$ & & -0.49$^{+0.42}_{-0.24}$ & 0.66$^{+0.12}_{-0.12}$          & 1.67           &    FB     \\
40020-01-03-00  & 26  & 4.69$^{+0.04}_{-0.05}$ & 6.20$^{+0.25}_{-0.25}$ & & ~1.36$^{+0.26}_{-0.27}$ & 1.24$^{+0.13}_{-0.15}$          & 1.80           &    HB     \\
\hline
\end{tabular}
\end{center}
\smallskip
\noindent
Note: bremsstrahlung + power law model used; 
\newline
\noindent
uncertainties are given at 90{\ts}{\%} confidence level for the derived
parameters of the model applied

\smallskip
\noindent
$^a$ Flux in 20--50{\ts}keV range  in units of 10$^{-9}$ ergs{\ts}cm$^{-2}${\ts}s$^{-1}$

\smallskip
\noindent
$^b$ Flux in 20--200{\ts}keV interval  in units of
                  10$^{-9}$ ergs{\ts}cm$^{-2}${\ts}s$^{-1}$
\end{table}

\begin{figure}
\begin{center}
\epsfig{file=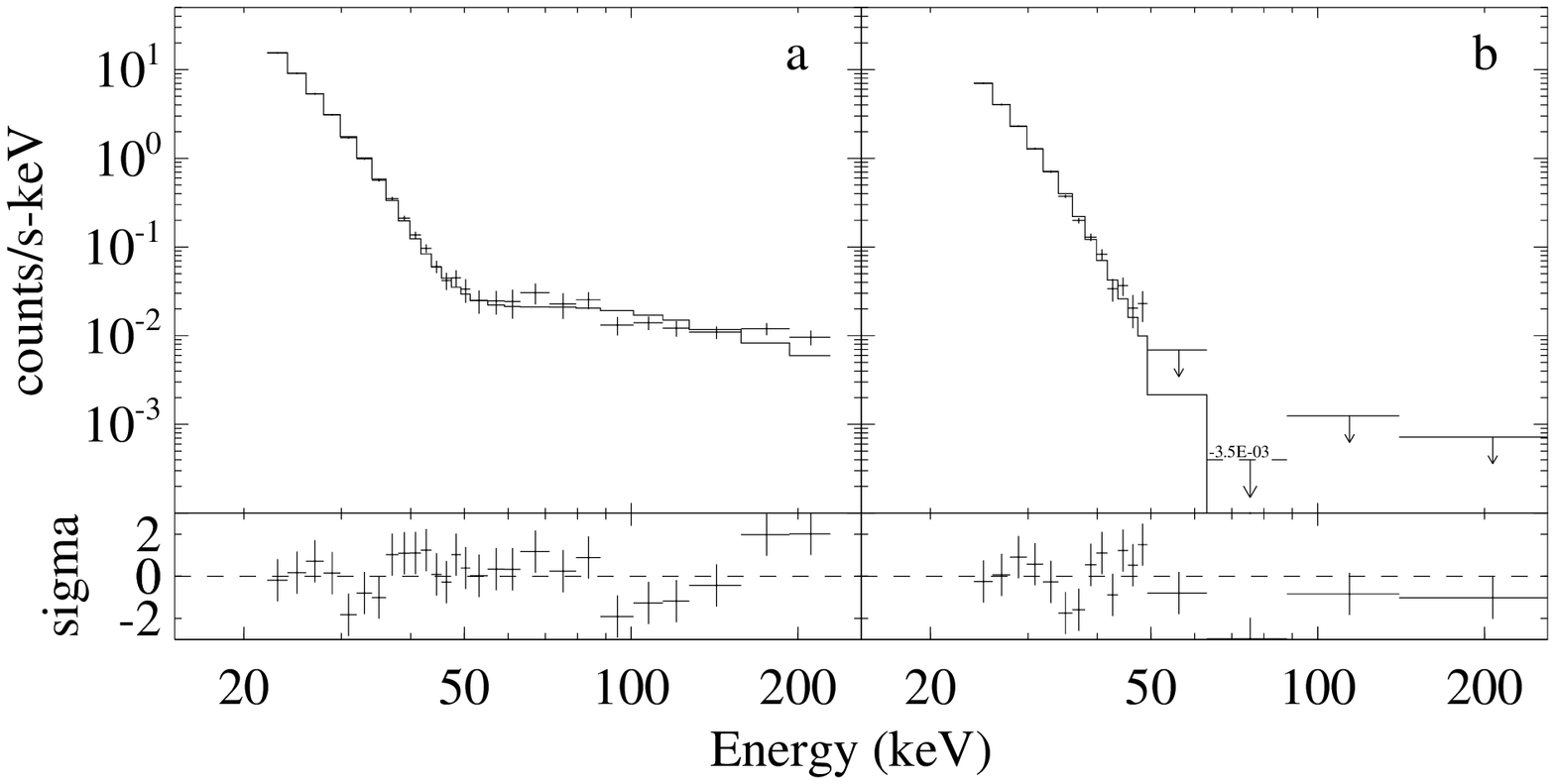, height=15cm, angle=90}
\caption{Spectrum resulting from a bremsstrahlung + power law
         fit to data subset 20053-01-02-00 (a), and 
         the result of a bremsstrahlung fit to subset
         30035-01-04-00 (b),
         showing the presence/absence of a hard X--ray tail in Sco X--1.
         Residuals are given in units of standard deviations (lower panels).  
         In (b) the upper limits are 2{\ts}$\sigma$ including the 60--80{\ts}keV
         bin which experienced a -{\ts}3{$\sigma$} residual.
         The ${\chi}^2_{\nu}$ are 1.27 and 1.62 respectively.}
\label{fig1}
\end{center}
\end{figure}

We observed a variation of a factor of $\sim${\ts}3 in the flux of the
non--thermal component in the 20--200{\ts}keV interval among our 8
detections, and, at least, an order of magnitude in overall variability,
taking into account our entire database. 

We were also able to study the time--scale of variations of
the hard X--ray component. As can be seen in Table
1, for observation 20 we detected a hard X--ray tail, 
which was not present 4 hours later in
observation 21. To our knowledge, this is the first time that such 
a short time--scale variability in the hard component is reported 
for Sco X--1.  

We also note that on all 3 occasions that the source was in the HB, a hard
X--ray tail was detected.

\section{DISCUSSION}

Taking 2.5{\ts}kpc as the lower limit to the distance to Sco X--1 
(Bradshaw, Fomalont, and Geldzahler, 1999), we found (using the
PCA data) Sco X--1 to be emitting near the Eddington level in the 
2--20{\ts}keV band (for a 1.4{\ts}M$_{\odot}$ neutron star). 
This is remarkably different from the atoll sources
(Barret et al., 2000), in which the luminosity in the same energy
range is below 10$^{37}${\ts}ergs{\ts}s$^{-1}$ when a hard tail is
detected. The 20--200{\ts}keV hard component luminosity
variation between our 8 detections is 
5.9--15.0{\ts}$\times${\ts}10$^{35}${\ts}ergs{\ts}s$^{-1}$, which is
comparable to the weaker atoll sources hard X--ray luminosities
reported by Barret, McClintock, and Grindlay (1996). 

We have observed the presence of a hard X--ray tail in each of the
3 branches on the {\Z} diagram. Although it is unclear
what is producing the hard X--ray tail in Sco X--1, as can be seen in 
Figure {\ref{fig2}}a, it appears that
the chance of observing a hard X--ray tail is greater if the thermal
component is brighter. This assumes that the thermal component
observed in HEXTE 20--50{\ts}keV band 
can be used to trace its $E${\ts}$<$ 20{\ts}keV
behavior. When we do observe a hard component, 
the power law index is correlated with the soft component temperature.
(Figure {\ref{fig2}}b). One other interesting and
suggestive relationship (Figure {\ref{fig2}}c) is that the hard
component flattens as the mass accretion rate
($\dot{M}$) increases, if we assume that the movement along
the {\Z} diagram (from the horizontal to the flaring branch) is the result of
the increasing variation of the $\dot{M}$ (see, e.g., van der Klis,
1995). Thus it is possible to speculate
that the mechanism responsible for the production of the hard component
and its shape is dependent on the $\dot{M}$, 
the 20--50{\ts}keV flux, and temperature of the soft thermal component. 
We note that while is straightforward to explain the derived power law
indices, and the hard X--ray emission, when the source was observed in the 
HB and in the FB in terms of synchrotron emission or Comptonization models, our
measured photon indices in the FB (consistent with 0) are not compatible
with the hard X--ray emission being produced by one of those two mechanisms.
 
\begin{figure}
\begin{center}
\epsfig{file=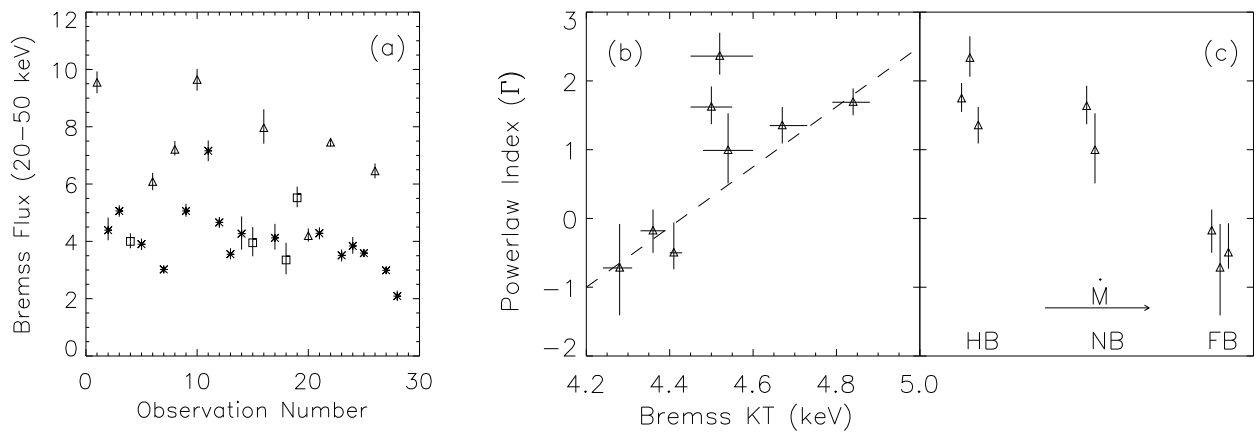, width=6cm, angle=90}
\caption{a) The thermal flux as measured by HEXTE (in units of
10$^{-9}${\ts}ergs{\ts}cm$^{-2}${\ts}s$^{-1}$) for our entire
database, which suggests that the chance of observing a hard X--ray tail
(triangles) is higher when the thermal component is brighter.
Squares are strong non--detections of a hard component (see Table 1)
and asterisks are treated as intermediate cases (see text). 
b) When a hard tail is observed, there is also some tendency for it to be
steeper as the temperature of the thermal component increases. c) Plotting 
the power--law observed against the branch in the {\Z} suggests that a
higher $\dot{\hbox{M}}$ makes the hard X--ray spectrum flatter (see
text for details). In (c), HB, NB and FB were arbitrarily spaced along the x
axis.}
\label{fig2}
\end{center}
\end{figure}

Our HEXTE observations 20--22 and 26 (see Table{\ts}1) overlap with BATSE/OSSE 
observations reported in Strickman and Barret (2000). It is interesting to
note that while HEXTE detected the presence of the hard X--ray tail in observations
20 and 22, BATSE/OSSE did not. Meanwhile, in observation 26 the hard X--ray tail
was detected both by HEXTE and BATSE/OSSE, with a spectrum extending up to
$\sim${\ts}400{\ts}keV. 

Our measured luminosities in the 2--20{\ts}keV ($L_x$) and 20--200{\ts}keV
energy intervals ($L_{hx}$) can be used to study the suggestion (Barret et al.,
2000) that such luminosities can be used to distinguish between neutron
star and black hole systems, with neutron stars systems emitting $L_{hx} \sim L_x$
when $L_x \ltsim L_{crit}$, with 
$L_{crit} = 1.5${\ts}$\times${\ts}10$^{37}${\ts}ergs{\ts}s$^{-1}$. 
Since Sco X--1 is emitting close to the Eddington level, our observations can 
not be used to test the hypothesis 
for the 2--20{\ts}keV luminosity (see details in Barret et al., 2000). 
Otherwise our observations of $L_{hx}$ are in agreement with the idea that  only
black hole binaries can have both $L_x$ and $L_{hx}$ above $L_{crit}$. 

\section{CONCLUSIONS}

In this paper, we provided clear evidence for variable hard X--ray emission from
the {\Z} source Sco X--1. We observed that the 20--200{\ts}keV
non--thermal component varied by at least an order of magnitude. 
We were able to track the movement of the source along the
{\Z} diagram and we concluded that the presence of the hard X--ray tail
is not confined to a particular position in such a diagram, which may
suggest that the mechanism responsible for the production of the hard
X--ray tail is not uniquely dependent on the inferred $\dot{M}$.
Otherwise we speculated that the appearance of a hard component in the
spectrum may be related to the brightness of the thermal component, and
that the shape of the X--ray tail may be correlated with  the temperature
of the thermal component. 

\section{ACKNOWLEDGMENTS}
This research has made use of data obtained through the High Energy
Astrophysics Science Archive Research Center Online Service, provided
by the NASA/Goddard Space Flight Center. F. D'Amico gratefully
acknowledges FAPESP/Brazil for financial support under grant
99/02352--2. This research was supported by NASA contract NAS5--30720.

\end{document}